\documentclass[prl,twocolumn,superscriptaddress,showpacs,preprintnumbers, amsmath,amssymb]{revtex4}
\usepackage{graphicx}
\def\be{\begin{equation}}
\def\ee{\end{equation}}
\def\ba{\begin{eqnarray}}
\def\ea{\end{eqnarray}}
\def\bw{\begin{widetext}}
\def\ew{\end{widetext}}

\begin{document}

\title{Transport through waveguides with surface disorder}

\author{M. Mart\'{\i}nez-Mares}
\affiliation{Departamento de F\'isica, Universidad Aut\'onoma 
Metropolitana-Iztapalapa, A. P. 55-534, 09340 M\'exico D. F., M\'exico}

\author{G. Akguc}
\altaffiliation{Present address: Cyprus International University, 
Nikosia, Cyprus}

\affiliation{Instituto de Ciencias F\'isicas, Universidad Nacional
Aut\'onoma de M\'exico, A. P. 48-3, 62210 Cuernavaca Mor., M\'exico}

\author{R.~A.~\surname{M\'endez-S\'anchez}}
\affiliation{Instituto de Ciencias F\'isicas, Universidad Nacional
Aut\'onoma de M\'exico, A. P. 48-3, 62210 Cuernavaca Mor., M\'exico}

\begin{abstract}
We show that the distribution of the conductance in 
quasi-one-dimensional systems with surface disorder is correctly 
described by the Dorokhov-Mello-Pereyra-Kumar equation if one includes 
direct processes in the scattering matrix $S$ through Poisson's kernel. 
Although our formulation is valid for any arbitrary number of channels, 
we present explicit calculations in the one channel case. Ours result is 
compared with solutions of the Schr\"odinger equation for waveguides 
with surface disorder calculated numerically using the $R$-matrix method.
\end{abstract}


\maketitle

The prediction of statistical properties of transport through disordered systems
is one of the fundamental problems in mesoscopic physics \cite{MelloKumar,Beenakker}.
The transmission coefficient, the dimensionless conductance if we are 
concerned with electronic devices, or in general any scattering quantity 
varies from sample to sample due to fluctuations on the microscopic configuration of disorder. Then, its distribution over an ensemble of systems macroscopically equivalent, but microscopically different, is of most interest \cite{MelloKumar}.
Transport through quasi-one dimensional waveguides (quantum wires 
in the electronic case) with {\em bulk} disorder has been 
tackled theoretically yielding to a Fokker-Planck equation, known as 
Dorokhov-Mello-Pereyra-Kumar (DMPK) equation \cite{D,MPK}. This equation gives 
the evolution of the probability density distribution of the transfer 
(scattering) matrix $M$ ($S$) when the length $L$ of the system increases. 
The parameter appearing as time scale in the DMPK equation is the diffusion time across the disordered region. This leads to a natural assumption, the {\em isotropic model} 
of uniformly distributed random phases for $M$ ($S$) \cite{MelloKumar}. 
In the jargon of nuclear physics, it means that 
there are not prompt responses that could arise from direct processes in 
the system \cite{Feshbach}.

An analytical solution of the DMPK equation, for any degree of disorder, 
was found in the absence of time-reversal symmetry ($\beta=2$ in Dyson's scheme 
\cite{Dyson}), by mapping the 
problem to a free-fermion model \cite{Beenakker1993,Beenakker1994}. 
In the presence of time-reversal invariance ($\beta=1$), solutions 
are known only in the localized \cite{Pichard} and metallic regimes \cite{Caselle}. 
Those solutions were recently checked with extensive numerical Monte Carlo 
simulations \cite{Garcia-Martin2001,Froufe-Perez2002}. It was also found in 
those references that waveguides with {\em surface} disorder are correctly 
described by the solutions of the DMPK equation in the localized regime.
Interestingly, the solutions of the DMPK equation does not fit with the numerical 
results for surface disorder in the crossover to-- nor in-- the metallic regime.

Up to now, a complete theory that include wires or waveguides with rough 
surfaces is missing. In this letter we show that systems with 
surface disorder are correctly described by the DMPK equation as the direct 
processes, due to short-direct trajectories connecting both sides of the 
waveguides, are taken into account. 
We introduce prompt responses in a {\em global} approach as explained below. 
The direct processes are quantified by the ensemble average $\langle S\rangle$ 
of the $S$-matrix, known in the literature as the {\em optical} $S$-matrix \cite{Lopez,Mello1985}. Our theoretical 
results are compared with numerical calculations based on 
the $R$-matrix theory \cite{Akguc} to solve the Schr\"odinger equation for the one-channel case.

\begin{figure}
\includegraphics[width=0.6\columnwidth]{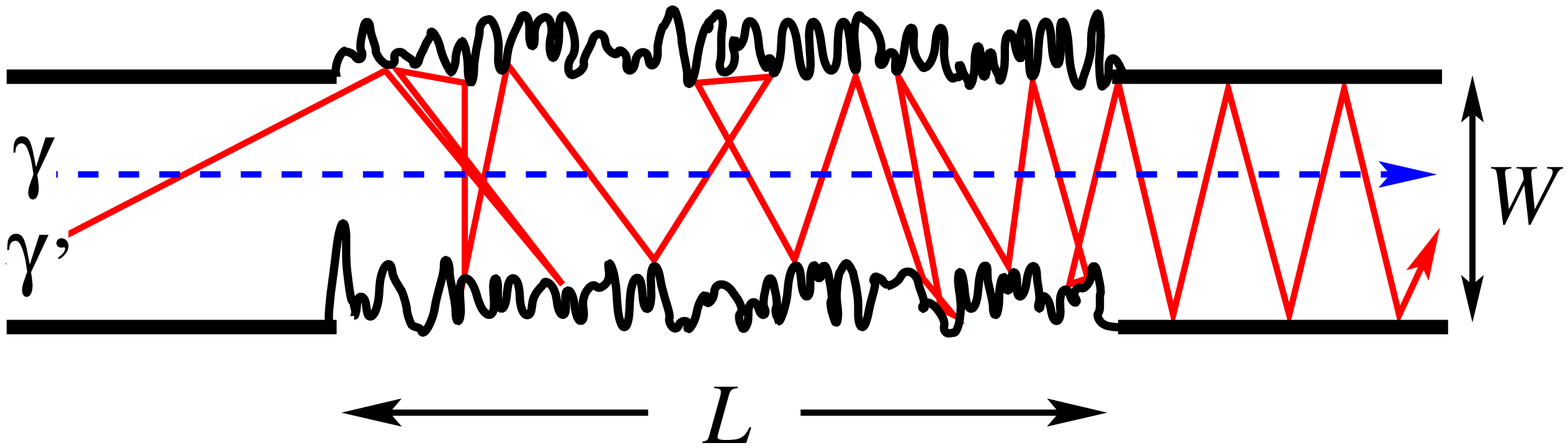} 
\caption{Sketch of a flat waveguide of width $W$ supporting 
$N$ channels with surface disorder in a region of length $L$. Two 
trajectories are also sketched. The trajectory $\gamma$ passes 
directly from one to the other side while trajectory $\gamma'$ 
passes randomly.}
\label{fig:guides}
\end{figure}

The waveguide with surface disorder is sketched in Fig.~\ref{fig:guides}. 
It consists in a flat waveguide of width $W$ that support $N$ open modes 
(or channels) with a  region of surface disorder of length $L$. The 
scattering problem is studied in terms of a $2N\times 2N$ scattering 
matrix $S$ which has the structure
\begin{equation}
S = \left( 
\begin{array}{cc}
r & t' \\ t & r' 
\end{array} 
\right),
\label{eq:S=rt} 
\end{equation}
where $r$ ($r'$) and $t$ ($t'$) are the reflection and transmission 
matrices for incidence on the left (right) of the disordered region. 
The dimensionless conductance $T=G/G_0$, where $G_0=2e^2/h$, is 
obtained from $S$ as $T = \rm{tr}(tt^{\dagger})$, according with 
Landauer's formula.

\begin{figure}
\includegraphics[width=0.8\columnwidth]{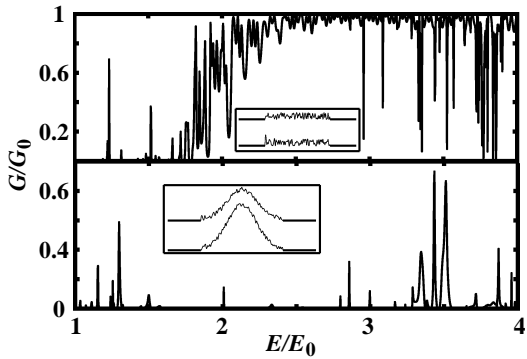} 
\caption{Dimensionless conductance $T=G/G_0$ as a function of 
$E/E_0$, where $E_0=\hbar^2\pi^2/2m$, of a waveguide with (a) surface 
disorder and direct transmission and (b) surface disorder but without 
direct transmission. The insets show the respective waveguides.
In (a) the conductance shows three regimes. In (b) the conductance
shows a localized phase only.}
\label{fig:conductance}
\end{figure}

\begin{figure}
\includegraphics[width=0.8\columnwidth]{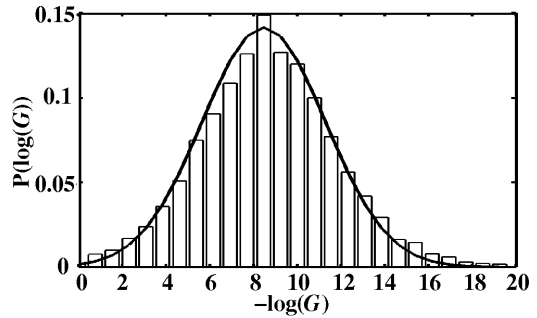} 
\caption{Distribution of the conductance (histogram) for an ensemble 
of 10 bent waveguides with surface disorder. It agrees with the 
solution of the DMPK equation in the localized regime (solid line).}
\label{fig:distribution}
\end{figure}

From Fig.~\ref{fig:guides}, it is clear that there could be direct 
transmission since there are trajectories, like $\gamma$, connecting both sides of the 
waveguide without any bounce in the rough surface although other trajectories like 
$\gamma'$ connect randomly both sides. In order to verify 
this statement we perform a numerical computation for the particular 
case $N=1$. We choose $W=1$; length $L=100$ is divided into 100 
pieces. To implement the surface disorder the ending point of each piece 
is a random number between 0 and a constant $\delta$ that 
measures the strength of the disorder. The ending points of each piece are connected with a spline interpolation to form a smooth disordered surface. The 
different realizations are done choosing different random 
displacements. We use a reaction matrix based method to solve the 
Schr\"odinger equation \cite{Akguc} in the scattering region with 
zero derivative at the leads. The basis states can be found using 
a metric for which scattering geometry transforms to a rectangular 
region. Finally, the scattering wavefunction is expanded in terms of 
those states.  The result for the dimensionless conductance $T$, 
for one realization of disorder, is shown in Fig.~\ref{fig:conductance}(a) 
as a function of the incident energy $E$ in units of the transverse energy 
$E_0=\hbar^2\pi^2/2m$. One channel is open for $E/E_0$ in the 
range 1 to 4. Within this range $T$ shows three different regions and not one 
as expected in quasi-one-dimensional problems with bulk disorder. For low 
$E/E_0$ (from 1 to 1.6 approximately) the conductance is small 
like in a ``localized'' regime. 
As shown in Refs.~\cite{Garcia-Martin2001,Froufe-Perez2002} 
the distribution of the conductance in the localized 
regime of waveguides with surface disorder agrees with the 
solution of the DMPK equation.
A ``metallic'' regime is obtained for energies between 2.5 and 4 for which
the conductance is close to 1. Finally, there is a transition region 
between the ``localized'' and ``metallic'' regimes. The existence of direct 
transmission in the waveguides with surface disorder is clear 
when a similar waveguide is bent in such a way that the direct 
trajectories are forbidden. In Fig.~\ref{fig:conductance}(b) the 
conductance of a bent waveguide with surface disorder is plotted. 
It shows only a localized phase. The distribution over an ensemble of 10 elements 
agrees with the DMPK solution in the localized regime (see Fig.~\ref{fig:distribution}). 
The appearance of only the localized phase in the bent waveguide 
yields strong evidence that the ``metallic'' region in 
Fig.~\ref{fig:conductance}(a) is due to direct processes. 

In the phenomenological model we include the direct processes in the 
solution of the 
DMPK equation. The $2N\times 2N$ scattering matrix $S$ of the system with 
$\langle S\rangle\neq 0$  is related to a $2N\times 2N$ scattering matrix 
$S_0$ with $\langle S_0\rangle=0$ through \cite{Hua,Friedman}
\begin{equation}
S_0 = \frac 1{t'_p} \left( S - \langle S\rangle_K \right) 
\frac 1{ ( \openone_{2N} - \langle S\rangle^{\dagger}_K S ) } t_p^\dagger.
\label{eq:transformation}
\end{equation}
where $\openone_{2N}$ is the unit matrix of dimension $2N$ and $t_p$ can be chosen as $t_p={t'_p}^{\dagger}$, where $t'_p$ is a $2N\times 2N$ matrix which satisfies 
$t'_p{t'_p}^{\dagger}= \openone_{2N}-\langle S\rangle_K\langle S\rangle^{\dagger}_K$ \cite{MelloKumar}. Here, $\langle S\rangle_{K}$ is the average of $S$ taken with Poisson's kernel, being the Jacobian of the transformation (\ref{eq:transformation}). The actual measured direct processes are quantified by $\langle S\rangle$ which is linearly related to $\langle S\rangle_{K}$ (see below). Then $S_0$ is a matrix that describes all the scattering process as $S$ with exception of the direct processes. It has a microscopic configuration of disorder that gives rise to $\langle S_0\rangle=0$. Note that $S$ is reduced to $S_0$ when $\langle S\rangle_{K}=0$ such that 
$\langle S\rangle=0$. Then the properties of the conductance in the 
``metallic'' regime of the waveguide with surface disorder can be obtained 
from the localized regime by adding the direct processes. This means 
that the scattering matrices $S$ and $S_0$ of the waveguide in the ``metallic'' 
and localized regimes, respectively, are related through 
Eq.~(\ref{eq:transformation}). 
Since the distribution of $S_0$ 
is given by the solution of the DMPK equation in the localized regime, 
the distribution of $S$ can be obtained from the 
distribution of $S_0$ taking into account the Jacobian of the 
transformation~(\ref{eq:transformation}).

As in Eq. (\ref{eq:S=rt}), $S_0$ has the structure 
\begin{equation}
S_0 = \left( 
\begin{array}{cc}
r_0 & t'_0 \\ t_0 & r'_0 
\end{array} 
\right),
\label{eq:S0par} 
\end{equation}
and both, $S$ and $S_0$, can be parameterized in a polar representation; 
for instance \cite{Hua,Baranger,MelloPichard,MPK} 
\begin{equation}
S_0 = \left[
\begin{array}{cc}
u_0^{(1)} & 0 \\ 0 & u_0^{(2)}
\end{array}
\right]
\left[
\begin{array}{cc}
-\sqrt{1-\tau_0} & \sqrt{\tau_0} \\ \sqrt{\tau_0} & \sqrt{1-\tau_0}
\end{array}
\right]
\left[
\begin{array}{cc}
u_0^{(3)} & 0 \\ 0 & u_0^{(4)}
\end{array}
\right],
\label{eq:S_0polar}
\end{equation}
where $\tau_0$ is a diagonal matrix whose elements are the eigenvalues 
$\{ {\tau_0}_n\in [0,1]\}$ of the matrix $t_0t_0^{\dagger}$ and $u_0^{(j)}$, 
$j=1,2,3,4$, are $N\times N$ unitary matrices for 
$\beta=2$, with the additional conditions $u_0^{(3)}={u_0^{(1)}}^T$ and $u_0^{(4)}={u_0^{(2)}}^T$ for $\beta=1$. In the isotropic model each $u_0^{(j)}$ is  distributed according to the invariant measure of the unitary group, $d\mu\left(u_0^{(j)}\right)$. The probability distribution of $S_0$ can be written as 
\begin{equation}
{dP_0}_{s_0}^{(\beta)}(S_0) = 
\frac{{P_0}_{s_0}^{(\beta)}(\{{\tau_0}_n\})}{p_{\beta}(\{{\tau_0}_n\})} 
d\mu_{\beta}(S_0), 
\label{eq:dPS0}
\end{equation}
where $s_0=L/\ell_0$ with $\ell_0$ the elastic mean free path without 
direct processes. ${P_0}_{s_0}^{(\beta)}(\{{\tau_0}_n\})$ is the solution 
of the DMPK equation \cite{Beenakker1993,Beenakker1994,Caselle} which depends on the elements of $\tau_0$ only, and 
\begin{equation}
d\mu_{\beta}(S_0) = p_{\beta}(\{{\tau_0}_n\})
\prod_{n=1}^N d{\tau_0}_n \prod_{j=1}^4 d\mu\left({u_0^{(j)}}\right)
\label{eq:measure1}
\end{equation}
(see Ref.~\cite{MelloKumar}) is the invariant measure of $S_0$ where 
\begin{equation}
p_{\beta}(\{{\tau_0}_n\}) = C_{\beta}
\prod_{a<b}{\tau_0}_a-{\tau_0}_b^{\beta} 
\prod_{c=1}^N {\tau_0}_c^{(\beta-2)/2}.
\label{eq:measure2}
\end{equation}

A similar parameterization as Eq.~(\ref{eq:S_0polar}) holds for $S$ and 
for its probability distribution we have 
\begin{equation}
dP_{s_0}^{(\beta)}(S) = \frac{P_{s_0}^{(\beta)}(S)}
{p_{\beta}(\{\tau_n\})} d\mu_{\beta}(S), 
\end{equation}
where $d\mu_{\beta}(S)$ has a similar expressions as Eqs.~(\ref{eq:measure1}) 
and~(\ref{eq:measure2}) suppressing the label ``0''. Now, 
$P_{s_0}^{(\beta)}(S)$ include the phases and it is obtained from the 
equation $dP_{s_0}^{(\beta)}(S)={dP_0}_{s_0}^{(\beta)}(S_0)$, which gives 
our main result, 
\begin{eqnarray}
P_{s_0}^{(\beta)}(S) & = & V_{\beta}^{-1}\, p_{\beta}(\{\tau_n\})
\frac{{P_0}_{s_0}^{(\beta)}(\{{\tau_0}_n(S)\})}
{ {p_{\beta}(\{{\tau_0}_n(S)\}) } } \nonumber \\ & \times &
\left[ 
\frac{ \det\left(\openone_{2N}-\langle S\rangle_K\langle S\rangle^{\dagger}_K\right)}
{\left|\det\left(\openone_{2N}-S\langle S\rangle^{\dagger}_K\right)\right|^2} 
\right]^{(2N\beta+2-\beta)/2}, \qquad
\label{P(S)}
\end{eqnarray}
where $V_{\beta}^{-1}$ is a normalization constant. The last term on 
the right hand side is the Jacobian of the 
transformation~(\ref{eq:transformation}),  
known as Poisson's kernel \cite{Hua,Friedman,Mello1985}. Here, we are postulating that this Jacobian is valid not only when $S_0$ is uniformly distributed but also for $S_0$ with uniformly distributed random phases. However, the average of $S$ that appears in the Jacobian is not the actual value of $\langle S\rangle$. In fact, it is satisfied the one-to-one correspondence $\langle S\rangle=\langle S\rangle_K+\langle S_f\rangle$, where $\langle S_f\rangle$ is the average of the fluctuating part of $S$, as can be easily 
seen inverting Eq.~(\ref{eq:transformation}) to write $S$ as $S=\langle S\rangle_K+S_f$. We have verified by numerically simulating DMPK $S_0$'s, and hence $S$ that, for a given value of $\langle S\rangle_K$, $\langle S_f\rangle$ coincides with $\langle S\rangle-\langle S\rangle_K$. Note that if $\langle S\rangle_K=0$ (at the same time $\langle S_f\rangle=0$), $S=S_0$ and $P_{s_0}^{(\beta)}(\{\tau_n\})$ reduces to 
${P_0}_{s_0}^{(\beta)}(\{{\tau_0}_n\})$. 
Finally, in Eq.~(\ref{P(S)}), it remains to write the ${\tau_0}_n$'s in 
terms of $S$ using Eq.~(\ref{eq:transformation}). 
As we mention before, 
we choose $t_p=t'_p$ once we calculate $t'_p$ numerically from the measured value of $\langle S\rangle$; in general it can be done by diagonalizing $t'_p{t'_p}^{\dagger}$. Then the distribution of the conductance $T$ can be obtained by multiple integration of Eq.~(\ref{P(S)}). An example is given below.

\paragraph{The case $\beta=1$ with $N=1$.}
In what follows we are concerned with the $\beta=1$ symmetry only and the index $\beta$ becomes irrelevant such that we can suppress it everywhere. In this case, $S_0$ is the $2\times 2$ matrix [see Eq.~(\ref{eq:S_0polar})]
\begin{equation}
S_0 = \left[ 
\begin{array}{cc}
-\sqrt{1-\tau_0}e^{2i\phi_0} & \sqrt{\tau_0} e^{i(\phi_0+\psi_0)} \\ 
\sqrt{\tau_0} e^{i(\phi_0+\psi_0)} & \sqrt{1-\tau_0}e^{2i\psi_0}
\end{array} 
\right],
\label{eq:polarN1}
\end{equation}
where $0\leq \phi_0,\psi_0<2\pi$ and $0\leq\tau_0\leq 1$. 
At the level of $S_0$, $\tau_0$ becomes the dimensionless conductance 
$T_0$, such that $\tau_0$ is replaced by $T_0$ everywhere. Eq.~(\ref{eq:measure1}) gives 
\begin{equation}
d\mu(S_0) = \frac{dT_0}{2\sqrt{T_0}}\, \frac{d\phi_0}{2\pi} \, \frac{d\psi_0}{2\pi}.
\label{eq:dm0}
\end{equation}
where we used that $p(T_0)=1/2\sqrt{T_0}$. 

\begin{figure}
\includegraphics[width=1.0\columnwidth]{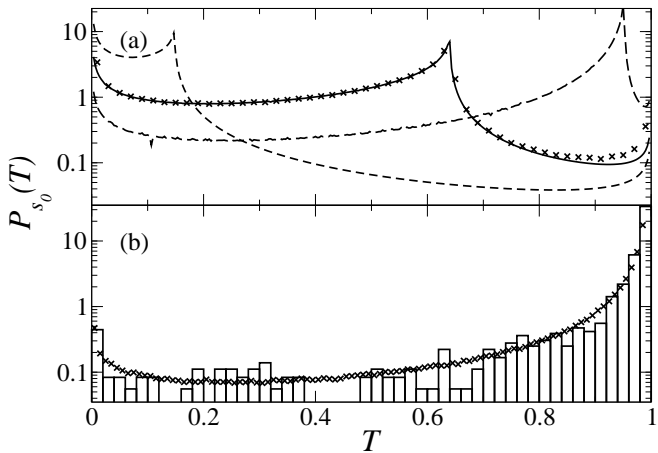} 
\caption{(a) Distribution of $T$ for the one channel case and $\beta=1$ for $s_0=10$, and intensities of the direct processes $w=0.2$ (dashed), 0.5 (continuous), 0.8 (long-dashed). The crosses 
are random matrix simulations ($10^6$ realizations). For clarity we 
present only the case $w=0.5$. The agreement is excellent. (b) Comparison between theory (random matrix simulations) and numerical solutions of Schr\"odinger equation as described in the text (histogram).
Here, $s_0=33.6$ and $w=0.93$. The agreement is excellent
but the experimental $\langle S\rangle$ does not fit the theoretical one.
}
\label{fig:theoryexp}
\end{figure}

As we mention above, the probability distribution of $S_0$ is given by the solution of the DMPK equation. Although the solution in any phase can be used, we use the one in the localized regime which is well known. In this phase, the variable $x_0$, defined by $T_0=1/{\cosh^2x_0}$, is Gaussian distributed \cite{Beenakker,Caselle}, 
\begin{equation}
{P_0}_{s_0}(x_0) = 
\frac 1{\sqrt{\pi s_0}} 
\exp \left[ -\frac 1{s_0} 
\left( x_0-\frac {s_0}2 
\right)^2 \right],
\label{eq:localized}
\end{equation}
with $s_0=-4\,\ln\langle T_0\rangle$. 
Equivalent expressions to Eqs.~(\ref{eq:polarN1}) and~(\ref{eq:dm0}) 
are valid for $S$ without the label ``0''. Those are used to write $T_0$ 
in terms of $S$ using Eq.~(\ref{eq:transformation}) for a given
$\langle S\rangle_K$. We will consider only the case 
$\langle S\rangle_K = w\,\sigma_x$, where $\sigma_x$ is one of 
the Pauli matrices and $w$ a complex number. 
For $t'_p$ we choose 
$t'_p=\sqrt{1-|w|^2} \mathrm{diag}(e^{i\theta_1} , e^{i\theta_2})$,
where $\theta_1$ and $\theta_2$ are arbitrary phases. From Eq.~(\ref{eq:transformation}) we get 
\begin{equation}
T_0 = \frac{ %(1-|v|^2)
\left|(1+|w|^2)\sqrt{T}e^{i\eta} - w^*e^{2i\eta} - w\right|^2}
{ \left| 1-2w^*\sqrt{T}e^{i\eta}+w^{*2}e^{2i\eta}\right|^2},
\label{eq:T0}
\end{equation}
where $\eta=\phi+\psi$. We note that the result is independent of the 
arbitrary 
phases $\theta_1$ and $\theta_2$. Finally, by direct substitution of 
$\langle S\rangle_K$, Eq.~(\ref{eq:localized}) and Eq.~(\ref{eq:T0}) into 
Eq.~(\ref{P(S)}), we obtain the result for $P_{s_0}(T,\phi,\psi)$, 
from which the distribution of $T$ is 
obtained by integration over the variables $\phi$ and $\psi$ in the range 0 to  $2\pi$. Since $\phi$ and $\psi$ always appear in the combination $\eta=\phi+\psi$ one integration can be done easily. The result is 
\begin{eqnarray}
&&\!\!\!\!\!\!\!\!\!\!\!P_{s_0}(T,\eta) = \frac 1{2\pi} \frac 1{\sqrt{\pi s_0}}
\exp\left[ -\frac 1{s_0}
\left( \cosh^{-1} \frac 1{\sqrt{T_0}} - \frac{s_0}2
\right)^2 \right] \nonumber
\\ & \!\!\!\!\!\!\!\!\!\!\times & 
\!\!\!\!\! \frac 1{2\sqrt{T}} \frac 1{\sqrt{T_0(1-T_0)}}\frac{\left(1-|w|^2\right)^{3}}
{\left| 1-2 w^*\sqrt{T}e^{i\eta}+w^{*2}e^{2i\eta}\right|^3}. \label{eq:distributiononechannel} 
\end{eqnarray}
The remaining integration over the variable $\eta$ can be done numerically. 
The results for $s_0=10$ and $w=0.2$, 0.5 and 0.8 are shown in 
Fig.~\ref{fig:theoryexp}(a), as well as one random matrix simulation only 
(for clarity of the figure). The simulation of $S$ is through 
Eq.~(\ref{eq:transformation}) with $S_0$ satisfying the 
DMPK equation in the localized phase. We observe an excellent 
agreement. As expected $P_{s_0}(T)$ moves from localized to 
metallic (from left to right) when $|w|$ take values from zero to one. 
The peak in the distribution of $T$ is probably reminiscent of 
the cut off observed in the numerical simulations with a higher number 
of channels of Ref.~\cite{Froufe-Perez2002}.

In Fig.~\ref{fig:theoryexp}(b), we compare the theory (simulations) 
with the numerical experiment of an ensemble of six realizations of 
flat waveguides with surface disorder, with the same parameters as 
those of Fig.~\ref{fig:conductance}. For $s_0=33.6$, we have an 
excellent agreement for $w=0.93$. The theory gives 
$\langle {S_f}_{11}\rangle=\langle {S_f}_{12}\rangle=0$, 
$\langle {S_f}_{21}\rangle=(1+i)\times 10^{-4}$, 
$\langle {S_f}_{22}\rangle=(-0.3+0.4\,i)\times 10^{-4}$, and 
it is satisfied that $\langle S\rangle = 0.93\sigma_x+\langle S_f\rangle$. However, the theoretical value of $\langle S\rangle$ does not 
coincide with the experimental one, which is a full matrix; for instance 
$\langle S_{12}\rangle_{\rm exp}=0.51582+0.06476\,i$. This is 
probably due to the simplicity of the model we used for  
$\langle S\rangle_K$, a full matrix being more realistic.

To conclude, we obtained the distribution of the conductance for 
waveguides with surface disorder. The direct transmission between 
both sides of the waveguide was taken into account in a global way. 
The results for the one channel case agree with numerical simulations, 
as well as with a numerical experiment made with R-matrix method. 
When the direct processes are increased the distribution moves from 
localized to metallic. Then, the direct processes could be the 
cause of the disagreement between the Dorokhov-Mello-Pereyra-Kumar 
equation and the numerical results recently obtained. 

This project was supported by DGAPA-UNAM under project IN118805. 
MMM received financial support from PROMEP-SEP through contract No.~34392. 
We thank P. A. Mello for useful discussions.

\end{document}